\documentclass{pj_sommerfeld}
\usepackage{amsmath,amsfonts,amssymb}
\usepackage{graphicx}
\usepackage{upgreek}
\usepackage{xcolor}
\usepackage{cite}
\usepackage[normalem]{ulem}
\usepackage{ragged2e}
\begin{document}
\setcounter{page}{1}
\pjheader{Vol.\ x, y--z, 2017}

\title[Progress In Electromagnetics Research, Vol. x, 2017]
{A Novel Asymptotic Solution to the Sommerfeld Radiation Problem: Analytic field expressions and the emergence of the Surface Waves}

\footnote{\it Received date}
\footnote{\hskip-0.12in*\, Corresponding author:~Panayiotis~Frangos (pfrangos@central.ntua.gr).}
\footnote{\hskip-0.12in\textsuperscript{1} Eurasian National University, 5, Munaitpassov Str., Astana, Kazakshtan.\newline\textsuperscript{2} School of Electrical and Computer Engineering, National Technical University of Athens, 9, Iroon Polytechniou Str., 157 73 Zografou, \newline\hspace*{3pt} Athens, Greece.} 
%\footnote{\hskip-0.12in\textsuperscript{2} School of Electrical and Computer Engineering, National Technical University of Athens, 9, Iroon Polytechniou Str., 157 73 Zografou, Athens, Greece.}

\author{Seil~Sautbekov\textsuperscript{1} , Sotiris~Bourgiotis\textsuperscript{2} , Ariadni~Chrysostomou\textsuperscript{2} and 
Panayiotis~Frangos\textsuperscript{*, 2}}

\runningauthor{Sautbekov, Bourgiotis, Chrysostomou and Frangos}

\tocauthor{Seil~Sautbekov, Sotiris~Bourgiotis , Ariadni~Chrysostomou 
and Panayiotis Frangos}

\begin{abstract}
The well-known ``Sommerfeld radiation problem'' of a small -Hertzian- vertical dipole above flat lossy ground is reconsidered. The problem is examined in the spectral domain, through which it is proved to yield relatively simple integral expressions for the received Electromagnetic (EM) field. Then, using the Saddle Point method, novel analytical expressions for the scattered EM field are obtained, including sliding observation angles. As a result, a closed form solution for the subject matter is provided. Also, the necessary conditions for the emergence of the so-called Surface Wave are discussed as well. A complete mathematical formulation is presented, with detailed derivations where necessary. 
\end{abstract}

%\mytableofcontents
%\tableofcontents

\setlength {\abovedisplayskip} {6pt plus 3.0pt minus 4.0pt}
\setlength {\belowdisplayskip} {6pt plus 3.0pt minus 4.0pt}

\

\section{Introduction}
\label{sec:introduction}

The ``Sommerfeld radiation problem'' is a well-known problem in the area of propagation of electromagnetic (EM) waves above flat and lossy ground with important applications in the area of wireless and mobile telecommunications \cite{1,2,3,4,5,6,7,8,9,10,11,12}. The original Sommerfeld solution to this problem is provided in the physical space by using the ``Hertz potentials'' and it does not end up with closed form analytical solutions. Subsequently, K. A. Norton \cite{13} focused  in the engineering application of the above problem and provided approximate solutions represented by rather long algebraic expressions.

In this paper, the authors advance on previous research work of theirs, concerning the solution of Sommerfeld\'{}s problem in the spectral domain. Namely, in \cite{14} the complete solution of the problem was given by means of an integral expression. In \cite{15,16} the Stationary Phase Method (SPM) \cite{17,18} was proposed and as a result closed-form analytic expressions were derived, for use in the high frequency regime and far way from the air-ground interface. However, that analysis did not consider the relative position of the stationary point to the integrand's singularities. The effect of that was that the analytic results failed to predict the well-known surface waves. Hence, in this paper the saddle point method is applied in such a manner that the EM field integral expression is transformed to a contour integral  of a special function, that possesses useful properties. The analysis is performed for the practical case of a conductive interface and ends up to novel formulas, providing clear physical understanding regarding the nature of the EM field, as well as the conditions under which the so called surface-wave appears. This resembles a cylindrical EM wave, propagating outwards, with respect to the dipole's horizontal distance, $\uprho$ and whose magnitude is exponentially decaying with respect to the altitude from the ground level, \emph{x} (see Fig.~\ref{fig:1} below). It is the asymptotic form that the general solution takes for sliding observation angles, otherwise (i.e. far away from the ground) converging to the usual form of an outgoing spherical wave.

The material is divided to six (6) sections, with Section \ref{sec:asym_expr} representing the core findings of this article, whereas Sections \ref{sec:prob_def}--\ref{sec:int_formul} describe the foundations and mathematical approach already been followed by this research team, as mentioned above. Important findings are summarized in Section \ref{sec:concl}. Derivations for most important statements and arguments are given in the appendices.
\section{Problem Definition}
\label{sec:prob_def}

\begin{figure}[h]
\centerline{\includegraphics[width=0.55\columnwidth,draft=false, height=0.30\columnwidth]{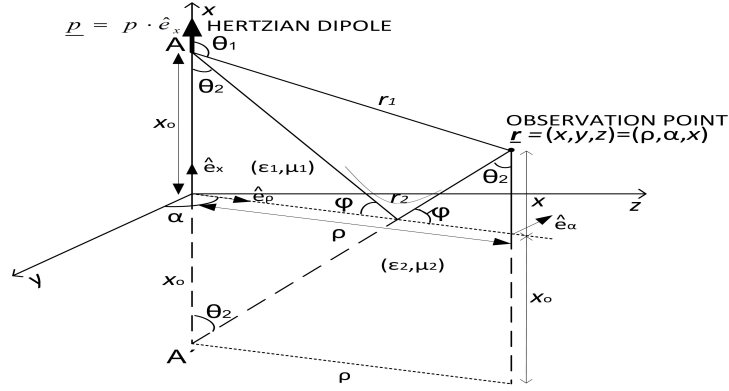}}
\caption{Hertzian Dipole above infinite, planar interface. Point A\'{} is the image of the source A with respect to the ground (yz-plane), r\textsubscript{1} is the distance between the source and the observation point, r\textsubscript{2} is the distance between the image of the source and the observation point, $\uptheta_{2}$ is the ``angle of incidence'' at the so-called ``specular point'', which is the point of intersection of the ground (yz-plane) with the line connecting the image point and the observation point, and finally, $\upphi = \pi/2 - \uptheta_{2}$ is the so-called ``grazing angle'' \cite{19}}
\label{fig:1}
\end{figure}

The problem geometry is provided in Fig.~\ref{fig:1}. A vertical small (Hertzian)  dipole, characterized by dipole moment $\underline{\dot{p}}=p\hat{e}_{x}, p\text{=const}$, is directed to the positive x axis, at altitude \textit{x\textsubscript{0}} above infinite, flat and lossy ground. The dipole radiates time-harmonic electromagnetic (EM) waves at angular frequency $\upomega = 2\pi f$ ($e^{-i\upomega t}$ time dependence is assumed). The relative complex permittivity of the ground is:~ $\upvarepsilon_{r}^{'} = \upvarepsilon^{'} / \upvarepsilon_{0} = \upvarepsilon_{r} + i \upsigma / \upomega \upvarepsilon_{0}$, where $\upsigma$ is the ground conductivity, $f$ the carrier frequency and $\upvarepsilon_{0} = 8.854\times10^{-12}$F/m is the permitivity in vacuum or air. The goal is to obtain closed form expressions for the received EM field at an arbitrary observation point above the ground level, namely at  point (x,y,z), shown in Fig.~\ref{fig:1}.

\vspace{-5mm}

\section{Problem Formulation}
\label{sec:prob_formul}

In general, the solution to the well - known Maxwell's equations can be written as \cite{14}:
\begin{equation} \label{eq:1}
\underline{E} = -\frac{i}{\upvarepsilon_{0}\upvarepsilon_{n}\upomega}
\left(\nabla\nabla\cdot + k_{0n}^2\right)\Uppsi_{n}\ast\underline{j}=
-\frac{i}{\upvarepsilon_{0}\upvarepsilon_{n}\upomega}F^{-1}\left\lbrace\left[k_{on}^2\underline{\tilde{J}} - (\underline{k}\cdot\underline{\tilde{J}})\underline{k}\right]\\\tilde{\Uppsi}_{n}\right\rbrace
\end{equation}

\begin{equation} \label{eq:2}
\underline{H} = -\left(\nabla\times\underline{j}\right)\ast\Uppsi_{n} = -iF^{-1}\left[\tilde{\Uppsi}_{n}\left(\underline{k}\times\underline{\tilde{J}}\right)\right]
\end{equation}

\noindent where $\underline{j}$ is the current density, $\ast$ is the convolution symbol through all coordinates, $\Uppsi_{n}, \tilde{\Uppsi}_{n}$ are the Green function and its Fourier transform respectively and $k_{0n}$, $n = \left\lbrace1,2\right\rbrace$ stand for the wave numbers of the first ($n = 1$) and second ($n = 2$) medium. Particularly, it holds true that \cite{14}:

\begin{equation} \label{eq:3}
\Uppsi_{n} = -\frac{e^{ik_{0n}r}}{4\uppi r},~~ \tilde{\Uppsi}_{n} = F\left\lbrace\Uppsi_{n}\right\rbrace = \left(k_{0n}^2 - k^2\right)^{-1}, ~~ k_{0n} = \upomega\sqrt{\upvarepsilon_{0}\upvarepsilon_{n}\upmu_{0}\upmu_{n}} 
\end{equation}

Due to the ``x-symmetry'' of the problem, the cylindrical coordinate system is most suitable for the decomposition of the various vector elements above; hence for example the wavevector is written as $\underline{k} = (k_{\uprho}, 0, k_{x})$.

\section{Integral Representations for the EM Field}
\label{sec:int_formul}

\subsection{Direct Field}
\label{subsec:dir_field_int}

The Hertzian source at point $\underline{r}_{0}=\left(x_0, 0, 0\right)$, shown in Fig.~\ref{fig:1}, corresponds to a current density of $\underline{j}^{e}= -i\upomega\underline{\dot{p}}\delta\left(\underline{r} - \underline{r_{0}}\right)$. Its Fourier transform is obviously: 

\begin{equation} \label{eq:4}
\underline{\tilde{J}}^{e} = F\left\lbrace\underline{j}^{e}\right\rbrace=\int_{-\infty}^{+\infty}\int_{-\infty}^{+\infty}\int_{-\infty}^{+\infty}~\underline{j}^{e}\cdot e^{-i\underline{k}\cdot\underline{r}}~~dxdydz = -i\upomega p\hat{e}_xe^{-ik_{x}x_{0}}
\end{equation}

\noindent for which the fact that $\underline{k}\cdot\underline{r} = k_{x }x+k_{\uprho} \uprho\cos \left(\alpha-k_{\alpha}\right)$ as well as the well known properties of the dirac function are considered.

Then, using (\ref{eq:1}) the direct or Line of Sight (LOS) field becomes:
\begin{align} \label{eq:5}
\underline{E}^{LOS}\hspace{-1mm} &= \hspace{-1mm}\frac{p}{\upvarepsilon_{0}\upvarepsilon_{1}}F^{-1}\hspace{-1mm}\left\lbrace \left(k_{x}\underline{k}-k_{01}^2\hat{e}_x\right)e^{-ik_{x}x_{0}}\hspace{-1mm}\cdot\hspace{-1mm}\large{\tilde{\Uppsi}_{1}}\right\rbrace \hspace{-1mm} = \hspace{-1mm} \frac{p}{\upvarepsilon_{0}\upvarepsilon_{1}}\frac{1}{(2\uppi)^{3}}\hspace{-1mm}\int_{0}^{\infty}\hspace{-3mm}\int_{0}^{2\pi}\hspace{-3mm}\int_{-\infty}^{\infty}\hspace{-3mm}\left( k_{x}\underline{k}-k_{01}^2\hat{e}_x\right)e^{-ik_{x}x_{0}}\hspace{-1mm}\cdot\hspace{-1mm}\large{\tilde{\Uppsi}_{1}}k_{\uprho} dk_{\uprho}dk_{\alpha}dk_{x}\hspace{-1mm}= \nonumber \\
&=\frac{p}{\upvarepsilon_{0}\upvarepsilon_{1}}\frac{1}{(2\uppi)^{3}}
\int_{0}^{\infty}\int_{0}^{2\pi}\int_{-\infty}^{+\infty}\hspace{-2mm}\left[\left(k_{x}^{2}-k_{01}^{2}\right)\hat{e}_x+k_{\uprho}k_{x}\hat{e}_{\uprho}\right]\large{\tilde{\Uppsi}_{1}}\cdot e^{ik_{x}\left(x-x_{0}\right)}e^{ik_{\uprho}\uprho\cos(k_{\alpha}-\alpha)}k_{\uprho} dk_{\uprho}dk_{\alpha}dk_{x}
\end{align}

\noindent Then in (\ref{eq:5}) and for the purposes of integrating over $k_{a}$, the integral representation for the Bessel function is used \cite{20}, namely $J_{0}\left(k_{\uprho}\uprho\right)=\dfrac{1}{2\uppi}\int_{0}^{2\pi}e^{ik_{\uprho}\uprho\cos k_{a}} dk_{a}$. Moreover, in \cite{1}, Sommerfeld showed that the $k_{\uprho}$ limits may be changed from $(0,\infty)$ to $(-\infty,\infty)$ by proving that for any function \textit{f}: $\int_{0}^{\infty}\hspace{-1mm}J_{0}\left(x\right)f\left(x\right)xdx\hspace{-1mm}=\hspace{-1mm}\dfrac{1}{2}\int_{-\infty}^{\infty}\text{H}_{~0}^{(1)}\left(x\right)f\left(|x|\right)xdx$, $\text{H}_{~0}^{(1)}$ being the Hankel Function of zero order and first kind\footnotemark
\footnotetext{Sommerfeld showed this by using properties of Bessel's and Hankel's functions: $J_{0}\left(z\right)=\dfrac{\text{H}_{~0}^{(1)}\left(z\right)+\text{H}_{~0}^{(2)}\left(z\right)}{2}, \text{H}_{~0}^{(1)}\left(ze^{i\pi}\right)=- \text{H}_{~0}^{(2)}\left(z\right)$}. As a result, and after performing basic algebraic calculations the following expression is obtained \cite{16}:

\begin{equation}\label{eq:6}
\underline{E}^{LOS} = \dfrac{p}{8\uppi^{2}\upvarepsilon_{0}\upvarepsilon_{1}}\int_{-\infty}^{+\infty}\int_{-\infty}^{+\infty}\left[\left(k_{x}^{2}-k_{01}^{2}\right)\hat{e}_{x}+|k_{\uprho}|k_{x}\hat{e}_{\uprho}\right]\large{\tilde{\Uppsi}_{1}}\cdot\text{H}_{~0}^{(1)}\left(k_{\uprho}\uprho\right)k_{\uprho}e^{ik_{x}(x-x_{0})}dk_{\uprho}dk_{x}
\end{equation}

Integration over $k_{x}$ can be performed through the use of the residue theory. Details, for the problem considered here, are found in \cite{16}. This results in the following one dimensional integral expression for the LOS Electric field:

\begin{align}\label{eq:7}
\underline{E}^{LOS}&=\dfrac{ip}{8\uppi\upvarepsilon_{0}\upvarepsilon_{1}}\int_{-\infty}^{+\infty}\left[\hat{e}_{x}\dfrac{k_{\uprho}^{2}}{\upkappa_{1}}-\hat{e}_{\uprho}|k_{\uprho}|\cdot\text{sgn}\left(x-x_{0}\right)\right]k_{\uprho}\text{H}_{~0}^{(1)}\left(k_{\uprho}\uprho\right)e^{i\upkappa_{1}|x-x_{0}|}~dk_{\uprho} = \nonumber
\\
&=-\dfrac{ip}{8\uppi\upvarepsilon_{0}\upvarepsilon_{1}}k_{01}\int_{-\infty}^{+\infty}\hat{e}_{\uptheta_1}\hspace{-1mm}\left(k_{\uprho}\right)\dfrac{k_{\uprho}|k_{\uprho}|}{\upkappa_{1}}\text{H}_{~0}^{(1)}\left(k_{\uprho}\uprho\right)e^{i\upkappa_{1}|x-x_{0}|}~dk_{\uprho}
\end{align}
\hspace{-1mm}\noindent with $\hat{e}_{\uptheta_1}\hspace{-1mm}\left(k_{\uprho}\right)\hspace{-1mm}=-\hat{e}_{x}\dfrac{|k_{\uprho}|}{k_{01}}+\hat{e}_{\uprho}\dfrac{\upkappa_{1}}{k_{01}}\cdot\text{sgn}\left(x-x_{0}\right)$ representing a unit vector and  $\upkappa_{1} = \sqrt{k_{01}^{2}-k_{\uprho}^{2}}$ . Similarly, the LOS Magnetic Field is given by:

\begin{equation}\label{eq:8}
\underline{H}^{LOS}=-\hat{e}_{a}\dfrac{i\upomega p}{8\uppi}\int_{-\infty}^{+\infty}k_{\uprho}\dfrac{|k_{\uprho}|}{\upkappa_{1}}\text{H}_{~0}^{(1)}\left(k_{\uprho}\uprho\right)e^{i\upkappa_{1}|x-x_{0}|}~dk_{\uprho}
\end{equation}

\subsection{Scattered Field}
\label{subsec:scat_field_int}

As known from elementary electromagnetic theory, an incident EM field that ``hits'' an interface enforces an ``ordered movement'' of the material's charge units; in other words, current flows are induced along the interface\footnotemark. These flows, can be modeled by the surface current densities $\underline{j}^{R}$ and $\underline{j}^{T}$, just above and below the interface level and act as the secondary sources for the reflected (x$>0$) and transmitted fields ($x<0$) respectively. For the vertical dipole case, considered hereby, the current densities exhibit a radial component only, ie $\underline{j}^{R}=\hat{e}_\uprho\cdot j^{R}$ and $\underline{j}^{T}=\hat{e}_\uprho\cdot j^{T}$

\footnotetext{this current flow is the compound result of the movement of the material's free charge units as well as the current that corresponds to the orientation of the material's dipole structures, commonly known as the ``displacement current''}

From this point onward, the same procedure of \ref{subsec:dir_field_int}, used for the LOS field, is followed here as well. The only difference is that the current sources $j^{R}$ and $j^{T}$ are unknown quantities and will need to be determined. This will be done by means of solving a boundary value problem at the interface level, as will be explained below. Hence, switching again to the frequency domain and applying (\ref{eq:1}) and (\ref{eq:2}) and after using the techniques described in \ref{subsec:dir_field_int} (eg. integrating over $k_{x}$ by means of the residue theory etc) the following integral expressions are obtained for the Reflected or Scattered (index ``\textit{R}'') and Transmitted (index ``\textit{T}'')  Electric and Magnetic field respectively [16]:

\begin{equation}\label{eq:9}
\underline{E}^{R}=-\dfrac{k_{01}}{8\uppi\upvarepsilon_{0}\upvarepsilon_{1}\upomega}\int_{-\infty}^{+\infty}\hat{e}_{\uptheta_2}\hspace{-1mm}\left(k_{\uprho}\right) k_{\uprho}\tilde{J}^R\cdot\text{H}_{~0}^{(1)}\left(k_{\uprho}\uprho\right)e^{i\upkappa_{1}x}~dk_{\uprho}
\end{equation}
\begin{equation}\label{eq:10}
\underline{H}^{R}=-\dfrac{\hat{e}_a}{8\uppi}\int_{-\infty}^{+\infty}k_{\uprho}\tilde{J}^R\cdot\text{H}_{~0}^{(1)}\left(k_{\uprho}\uprho\right)e^{i\upkappa_{1}x}~dk_{\uprho}
\end{equation} 

\begin{equation}\label{eq:11}
\underline{E}^{T}=-\dfrac{k_{02}}{8\uppi\upvarepsilon_{0}\upvarepsilon_{2}\upomega}\int_{-\infty}^{+\infty}\hat{e}'_{\uptheta_2}\hspace{-1mm}\left(k_{\uprho}\right) k_{\uprho}\tilde{J}^T\cdot\text{H}_{~0}^{(1)}\left(k_{\uprho}\uprho\right)e^{-i\upkappa_{2}x}~dk_{\uprho} 
\end{equation}

\begin{equation}\label{eq:12}
\underline{H}^{T}=\dfrac{\hat{e}_a}{8\uppi}\int_{-\infty}^{+\infty}k_{\uprho}\tilde{J}^T\cdot\text{H}_{~0}^{(1)}\left(k_{\uprho}\uprho\right)e^{-i\upkappa_{2}x}~dk_{\uprho}, ~~~~\upkappa_{2} = \sqrt{k_{02}^{2}-k_{\uprho}^{2}}
\end{equation} 

\noindent In (\ref{eq:9}) and (\ref{eq:11}), a notation similar to that of (\ref{eq:7}), for the LOS field, is used. Namely, it holds that $\hat{e}_{\uptheta_2}\hspace{-1mm}\left(k_\uprho\right)=1/k_{01}\cdot\left(\upkappa_{1}\hat{e}_{\uprho} - |k_{\uprho}|\hat{e}_{x}\right)$ and $\hat{e}'_{\uptheta_2}\hspace{-1mm}\left(k_\uprho\right)=1/k_{02}\cdot\left(\upkappa_{2}\hat{e}_{\uprho} + |k_{\uprho}|\hat{e}_{x}\right)$, each representing a unit vector with respect to dummy variable $k_\uprho$, applicable for the reflected and transmitted electric fields respectively.

The boundary value problem, mentioned above, dictates that at the planar interface ($x=0$), that is at the ground level, the tangential components of the EM field should be continuous. In other words it should be both valid:

\begin{equation}\label{eq:13}
E_{\uprho}^{LOS}|_{x=0} + E_{\uprho}^{R} |_{x=0} = E_{\uprho}^{T}|_{x=0}~~~~ H_{a}^{LOS}|_{x=0} + H_{a}^{R}|_{x=0} = H_{a}^{T}|_{x=0}
\end{equation}

Substituting (\ref{eq:9}) - (\ref{eq:12}) to (\ref{eq:13}), the following system of algebraic equations and its associated solution for $\tilde{J^R}$ and $\tilde{J^T}$ is obtained:

\begin{minipage}{0.45\textwidth}
\begin{equation*}
\begin{cases}
&i\upomega p|k_{\uprho}|e^{i\upkappa_{1}x_{0}} + \upkappa_{1}\tilde{J}^R = -\upkappa_{1}\tilde{J}^T \\
\\
&-i\upomega p|k_{\uprho}|e^{i\upkappa_{1}x_{0}} + \upkappa_{1}\tilde{J}^R = \dfrac{\upvarepsilon_{1}}{\upvarepsilon_{2}}\upkappa_{2}\tilde{J}^T\\
\end{cases}
\end{equation*}
\end{minipage}
$\quad \hspace{-9mm}\Longrightarrow $\quad 
\begin{minipage}{0.47\textwidth}
\begin{equation} \label{eq:14}
\begin{cases}
& \tilde{J}^R = i\upomega p|k_{\uprho}|e^{i\upkappa_{1}x_{0}}\dfrac{\upvarepsilon_{2}\upkappa_{1}-\upvarepsilon_{1}\upkappa_{2}}{\upkappa_{1}\left(\upvarepsilon_{2}\upkappa_{1}+\upvarepsilon_{1}\upkappa_{2}\right)}\\
\\
&\tilde{J}^T = -i\upomega p|k_{\uprho}|e^{i\upkappa_{1}x_{0}}\dfrac{2\upvarepsilon_{2}}{\upvarepsilon_{2}\upkappa_{1}+\upvarepsilon_{1}\upkappa_{2}}\\
\end{cases}
\end{equation}
\end{minipage}

\vspace{\baselineskip}
Obviously, by simple substitution of (\ref{eq:14}) to (\ref{eq:9}) - (\ref{eq:12}), the complete formulas for the reflected and transmitted EM field in their integral form are obtained.

\section{Asymptotic Analytic Expressions for the EM Field}
\label{sec:asym_expr}

\subsection{Direct Field}
\label{subsec:dir_field_anal}

Now, refer to (\ref{eq:7}) and set: $k_{\uprho}=k_{01}\sin\upxi\Rightarrow dk_{\uprho} = k_{01}\cos\upxi d{\upxi}$, $\upkappa_{1}\hspace{-1mm} =\hspace{-1mm} k_{01}\cos{\upxi}$. Also, the large argument approximation for the Hankel function is considered \cite{20}, namely: $\text{H}_{~0}^{(1)}\left(k_{\uprho}\uprho\right)\simeq\sqrt{\dfrac{-2i}{\uppi k_{\uprho}\uprho}}e^{ik_{\uprho}\rho},~k_{\uprho}\uprho\gg 1$. These transform (\ref{eq:7}) to:\footnotemark 

\footnotetext{The transformation of $\left(-\infty, +\infty\right)$ to the $ S_{z}$  contour (see Fig.~\ref{fig:2}) is explained in Appendix \ref{App:A} }

\begin{equation}\label{eq:15}
\hspace{-3pt}\underline{E}^{LOS}\hspace{-4pt}=-\dfrac{ip}{8\uppi\upvarepsilon_{0}\upvarepsilon_{1}}\sqrt{\dfrac{-2i}{\uppi k_{01}\uprho}}k_{01}^{3}\hspace{-4pt}\int\limits_{S_{z}}\hat{e}_{\uptheta_1}\hspace{-1mm}\left(\upxi\right)~\sin^{^{\frac{3}{2}}}\hspace{-1mm}\upxi\cdot e^{ik_{01}\left(\uprho \sin\upxi + |x-x_{0}|\cos \upxi\right)} d\upxi
\end{equation}

Referring to Fig.~\ref{fig:1}, it is obvious that $\uprho = r_{1} \sin\uptheta_{1}$ and $|x-x_{0}|=r_{1}|\cos\uptheta_{1}|$. Then, taking for example, the case where $x>x_{0}$ (the same results occur for $x<x_{0}$), i.e. when the observation point is higher than the transmitting dipole, it holds that:

\begin{equation}\label{eq:16}
\underline{E}^{LOS}=-\dfrac{ip}{8\uppi\upvarepsilon_{0}\upvarepsilon_{1}}\sqrt{\dfrac{-2i}{\uppi k_{01}\uprho}}k_{01}^{3}\int\limits_{S_{z}}\hat{e}_{\uptheta_1}\hspace{-1mm}\left(\upxi\right)~\sin^{^{\frac{3}{2}}}\hspace{-1mm}\upxi\cdot e^{ik_{01}r_{1}\cos\left(\upxi - \uptheta_{1}\right)}~d\upxi
\end{equation}

Note that under the aforementioned transformation, $k_{\uprho}=k_{01}\sin\upxi$, the unit vector with respect to  $k_{\uprho}$, defined in (\ref{eq:7}), is transformed to $\hat{e}_{\uptheta_1}\left(\upxi\right) = \cos\upxi\cdot\hat{e}_{\uprho}-|\sin\upxi|\cdot\hat{e}_{x}$. Consequently, $\hat{e}_{\uptheta_1}\left(\upxi\right)$ may be regarded as a pseudo unit vector along the complex elevation angle $\upxi$\footnote{\label{pseudo_vector}It is called a pseudo unit vector because it is associated with the complex angle $\upxi$ and hence has no physical meaning. However, at $\upxi=\uptheta_{1}$ it becomes equal to the geometric unit vector (since for the spherical coordinate system with O$\equiv$A: $\hat{e}_{\uptheta_{1}} = \hat{e}_{\uprho}\cos\uptheta_{1} - \hat{e}_{x}\sin\uptheta_{1}$)}. The previous concept is graphically depicted in Fig.~\ref{fig:2}, where the extended local spherical coordinate systems for the dipole source and its associated image are drawn. The term ``extended'' is used to denote the complex nature of the elevation angle $\upxi$ along the $S_z$ contour of (\ref{eq:15}), (\ref{eq:16}), above, or of (\ref{eq:19}) - (\ref{eq:21}), below.

%\begin{displaymath}
%\hat{e}_{\uptheta_1}\left(\upxi\right) = \frac{\cos\upxi |\sin\upxi|}{\sin\upxi}\textnormal{sgn}\left(x-x_0\right)\cdot\hat{e}_{\uprho}-\sin\upxi\cdot\hat{e}_{x}=\cos\upxi\cdot\hat{e}_{\uprho}-\sin\upxi\cdot\hat{e}_{x}
%\end{displaymath}

The integral expression of (\ref{eq:16}) can be easily evaluated with the use of the Saddle Point method \cite{18}, on the precondition that parameter ``$k_{01}\cdot r_{1}$'' of the phase factor of (\ref{eq:16}), is sufficiently large. Actually, since the integrand does not present any singularity, the result is equivalent to that given by the Stationary Phase Method \cite{17,18}, given in \cite{21,22}:
\begin{equation}\label{eq:17}
\underline{E}^{LOS}\simeq -\dfrac{pk_{01}^{2}}{4\uppi\upvarepsilon_{0}\upvarepsilon_{1}r_{1}}\sin\uptheta_{1} e^{ik_{01}r_{1}}\left(\hat{e}_{\uprho}\cos\uptheta_{1} - \hat{e}_{x}\sin\uptheta_{1}\right) = -\dfrac{pk_{01}^{2}}{4\uppi\upvarepsilon_{0}\upvarepsilon_{1}r_{1}}\sin\uptheta_{1} e^{ik_{01}r_{1}}\hat{e}_{\uptheta_{1}} 
\end{equation}

Starting from (\ref{eq:2}), the analysis for the direct magnetic field is totally similar. The asymptotic expression for the far field thus becomes:

\begin{equation}\label{eq:18}
\underline{H}^{LOS} = -\hat{e}_a\dfrac{i\upomega p}{8\uppi}\int_{-\infty}^{+\infty}\dfrac{k_{\uprho}|k_{\uprho}|}{\upkappa_{1}}\cdot\text{H}_{~0}^{(1)}\left(k_{\uprho}\uprho\right)e^{i\upkappa_{1}|x-x_{0}|}~dk_{\uprho} \simeq \hat{e}_a \dfrac{pk_{01}^{2}\sin\uptheta_{1} e^{ik_{01}r_{1}}}{4\uppi\sqrt{\upvarepsilon_{0}\upvarepsilon_{1}\upmu_{0}\upmu_{1}}\cdot r_{1}}= -\hat{e}_a\dfrac{1}{Z_{1}}\vert\underline{E}^{LOS}\vert 
\end{equation}

\noindent with $Z_{1} = \sqrt{\upmu_{0}\upmu_{1}/\upvarepsilon_{0}\upvarepsilon_{1}}$, the wave impedance of medium 1 (air)\footnotemark. Expressions (\ref{eq:17}) and (\ref{eq:18}) reflect the well known, in the literature,  far field expressions of a Hertzian dipole source \cite{17,19} and as such validate the spectral domain presented so far.

\footnotetext{Similarly, the wave impedance of medium 2 (ground) is given by: $Z_{2} = \sqrt{\upmu_{0}\upmu_{2}/\upvarepsilon_{0}\upvarepsilon_{2}}$}

\begin{figure}[h]
\centerline{\includegraphics[width=0.80\columnwidth,draft=false]{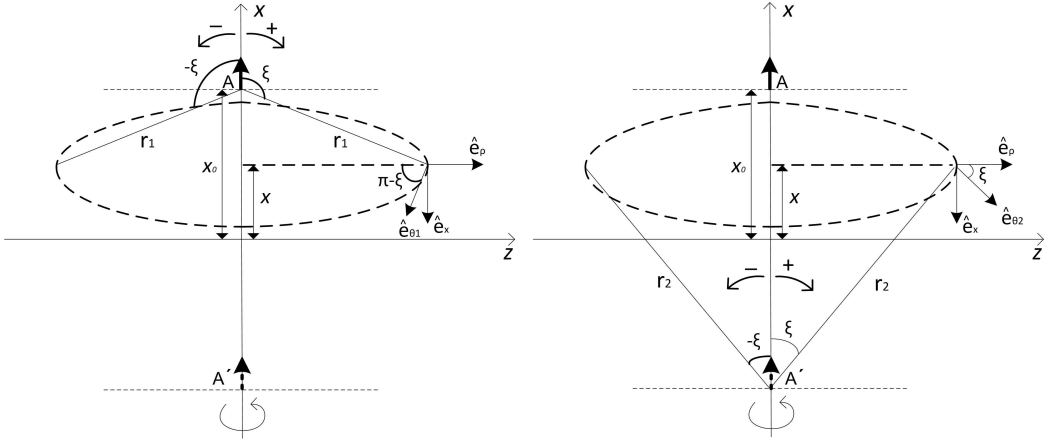}}
\caption{The ``extended'' local spherical coordinate system of the dipole source (left) and the dipole image (right). Elevation angle $\upxi$ is complex in nature, however only the range $\upxi\in\left[-\uppi, \uppi\right]$ may be graphically depicted  (in a normal sperical coordinate system, however, $\upxi\in\left[0, \uppi\right]$). In the figure, a clockwise rotation from the positive x-direction yields a positive elevation angle (+), whilst an anticlockwise turn generates a negative elevation angle (-).}
\label{fig:2}
\end{figure}

\subsection{Scattered Field}
\label{subsec:scat_field_anal}

Similarly to \ref{subsec:dir_field_anal}, the variable transformation $k_{\uprho}=k_{01}\sin \upxi_2$ is applied to (\ref{eq:9}). With the large argument approximation for the Hankel function, also considered, this results in the following integral expression for the reflected field:
\begin{equation}\label{eq:19}
\underline{E}^{R}=-\dfrac{ipk_{01}^{3}}{8\uppi\upvarepsilon_{0}\upvarepsilon_{1}}\sqrt{\dfrac{-2i}{\uppi k_{01}\uprho}}\int\limits_{S_{z}}\hat{e}_{\uptheta_2}\hspace{-1mm}\left(\upxi\right)~\sin^{^{\frac{3}{2}}}\hspace{-1mm}\upxi\cdot R_{\parallel}\left(\upxi\right) e^{ik_{01}\left[\uprho \sin\upxi + \left(x+x_{0}\right)\cos\upxi\right]} ~d\upxi
\end{equation}

\noindent with {$R_{\parallel}\left(\upxi\right)$} being the Fresnel reflection coefficient, given by:

\begin{equation}\label{eq:20}
R_{\parallel}\left(\upxi\right) = \frac{k_{01}\upvarepsilon_{2}\cos\upxi - \upvarepsilon_{1}\sqrt{k_{02}^{2} - k_{01}^{2}\sin^{2}\upxi}}{k_{01}\upvarepsilon_{2}\cos\upxi + \upvarepsilon_{1}\sqrt{k_{02}^{2} - k_{01}^{2}\sin^{2}\upxi}} = \frac{k_{02}Z_{1}\cos\upxi - Z_{2}\sqrt{k_{02}^{2} - k_{01}^{2}\sin^{2}\upxi}}{k_{02}Z_{1}\cos\upxi + Z_{2}\sqrt{k_{02}^{2} - k_{01}^{2}\sin^{2}\upxi}}
\end{equation}

Then, with a glance back to the geometry of Fig.~\ref{fig:1} it is easy to identify that $\uprho = r_{2}\sin\uptheta_{2}$ and $x+x_{0}=r_{2}\cos\uptheta_{2}$. Substituting to (\ref{eq:19}) the expression for the scattered field becomes:

\begin{equation}\label{eq:21}
\underline{E}^{R}=-\dfrac{ipk_{01}^{3}}{8\uppi\upvarepsilon_{0}\upvarepsilon_{1}}\sqrt{\dfrac{-2i}{\uppi k_{01}\uprho}}\int\limits_{S_{z}}\hat{e}_{\uptheta_{2}}\hspace{-1mm}\left(\upxi\right)~\sin^{^{\frac{3}{2}}}\hspace{-1mm}\upxi\cdot R_{\parallel}\left(\upxi\right) e^{ik_{01}r_{2}\cos\left(\upxi-\uptheta_{2}\right)} ~d\upxi
\end{equation}

\vspace{-1mm}\noindent where $\displaystyle\hat{e}_{\uptheta_{2}}\left(\upxi\right)=\left(\hat{e}_{\uprho}\cos\upxi - \hat{e}_{x}|\sin\upxi|\right)$, a pseudo unit vector, along the complex elevation angle $\upxi$, of a spherical coordinate system, this time with its origin located at the dipole's image point, A$'$ of Fig.~\ref{fig:2}\hspace{-1mm}$~^{\parallel}$.

The presence of $R_{\parallel}\left(\upxi\right)$, imposes a pole to the integrand of (\ref{eq:21}), found by solving $1/R_{\parallel}\left(\upxi\right) = 0$:

\begin{equation}\label{eq:22}
\cos\upxi_{\text{p}} = -\frac{\upvarepsilon_{1}}{k_{01}}\sqrt{\frac{k_{02}^{2}-k_{01}^{2}}{\upvarepsilon_{2}^{2}-\upvarepsilon_{1}^{2}}}=-\sqrt{\frac{\upvarepsilon_{1}\left(\upvarepsilon_{2}\upmu_{2}-\upvarepsilon_{1}\upmu_{1}\right)}{\upmu_{1}\left(\upvarepsilon_{2}^{2}-\upvarepsilon_{1}^{2}\right)}}~\stackrel{\upmu_{1}\simeq\upmu_{2}}{=} -\sqrt{\frac{\upvarepsilon_{1}}{\upvarepsilon_{1}+\upvarepsilon_{2}}}
\end{equation}

\noindent where the rightmost side of (\ref{eq:22}) considers the ordinary case where $\upmu_{1}=\upmu_{2}$. Moreover, for the usual case of $\upsigma\gg \upomega\upvarepsilon_{0}$, the complex (effective) relative permitivity of medium 2, given by $\dot{\upvarepsilon}_{2} = \upvarepsilon_{2} + i\frac{\upsigma}{\upomega\upvarepsilon_{0}}$, is a large volume quantity. Note that in (\ref{eq:22}) and in the whole analysis, given so far, $\upvarepsilon_{2}$ actually refers to this complex relative permitivity, ie $\upvarepsilon_{2}\equiv\dot{\upvarepsilon}_{2}$\footnotemark . As a result (\ref{eq:22}) yields a small (complex) number. Then following the reasoning of Appendix \ref{App:B} the pole can be estimated by:

\begin{equation}\label{eq:23}
\upxi_{\text{p}}\simeq \frac{\uppi}{2}+\sqrt{\frac{\upomega\upvarepsilon_{0}\upvarepsilon_{1}}{2\upsigma}}\left\lbrace 1+\frac{\upomega\upvarepsilon_{0}\left(\upvarepsilon_{1}+\upvarepsilon_{2}\right)}{2\upsigma}-i\left[1-\frac{\upomega\upvarepsilon_{0}\left(\upvarepsilon_{1}+\upvarepsilon_{2}\right)}{2\upsigma}\right]\right\rbrace
\end{equation}
\noindent in which case, of course, $\upvarepsilon_{2}$ now refers to the usual, real (relative) permitivity part of medium 2 (ground).

\begin{figure}[h]
\centerline{\includegraphics[width=0.65\columnwidth,draft=false]{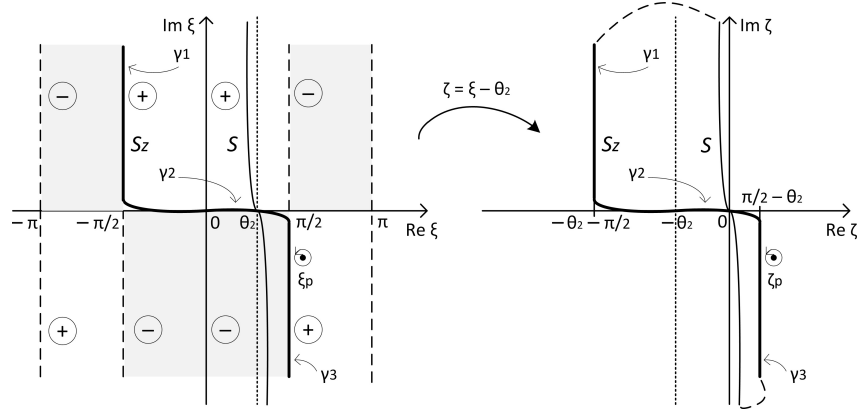}}
\caption{The contour of integration: a)$S_{z}$: original contour for $\underline{E}^{R}$ in the complex $\upxi$-plane (left plot) and its $\upzeta$-plane mapping (right plot)  b)$S$: ``Etalon integral'' contour in the $\upzeta$-plane (right plot) and its $\upxi$-plane mapping (left plot), c)$\upxi_{p}$: relative position of the pole in the $\upxi$-plane, d) $\upzeta_{p}$: relative position of the pole in the $\upzeta$-plane}
\label{fig:3}
\end{figure}

\footnotetext{see respective definitions and notation used in Section \ref{sec:prob_def} and Section \ref{sec:prob_formul}}
\enlargethispage{2\baselineskip}
The contour of integration, $S_{z}$ together with the position of the pole, $\upxi_{\text{p}}$ are shown in Fig.~\ref{fig:3}. For $\upsigma\gg\upomega\upvarepsilon_{0}$, $\upxi_{\text{p}}$ is nearby $S_{z}$, particularly to the right of the $\upgamma_{3}$ segment of it. As a result, directly evaluating (\ref{eq:19}) by the method of saddle points is likely to induce significant errors, since the accuracy of the method depends on the relative position of the pole to the saddle point \cite{17,18}. This was actually revealed in \cite{22,23} in which the evaluation of (\ref{eq:19}) resulted in a single term, corresponding to the space wave component of the field only, thus missing to describe the surface wave behavior of it. Even, more, at sliding observation angles, the method failed completely to estimate field values, a consequence of the known fact that the space wave component at such circumstances almost vanishes, for the reflection coefficient being almost equal to minus one \cite{19}.

%\paragraph{•}
To increase the accuracy of integration the so called ``Etalon Integral''\footnotemark\cite{24,25,26,27} given by:
\footnotetext{The term ``Etalon Integral'' is used here to denote a kind of ``standard'', or ``reference integral''.}

\begin{equation}\label{eq:24}
X\left(k,\alpha\right)=\frac{1}{4\uppi i}\int_{S}\frac{e^{ik\left(\cos\upzeta-\cos\alpha\right)}}{\sin\frac{\upzeta+\alpha}{2}}~d\upzeta
\end{equation}
\noindent is used, which reduces the problem related to the vicinity of the pole point to the saddle point. The integration contour for $X\left(k,\alpha\right)$, \textit{S}, is shown in the right hand side of Fig.~\ref{fig:3} and passes downwards along the imaginary axis, deviating slightly from it to ensure convergence.

Special function $X\left(k,\alpha\right)$ has interesting properties, useful for the purposes of the problem considered here. It is expressed in terms of the Fresnel Integrals or probability integrals, which are well known special functions with numerous applications (see also Appendix \ref{App:C}):

\begin{equation}\label{eq:25}
X\left(k,\alpha\right)=\frac{e^{-i\frac{\uppi}{4}}}{\sqrt{2\uppi}}\int_{\infty\sin\frac{\alpha}{2}}^{2\sqrt{k}\sin\frac{\alpha}{2}}\large{e^{\frac{it^{2}}{2}}~dt}= -\frac{1}{2}\text{sgn}\left(\alpha\right)\text{erfc}\left[\text{sgn}\left(\alpha\right)\sqrt{-2ik}\sin\frac{\alpha}{2}\right]
\end{equation} 

\noindent Also, asymptotic formulas for large values of the upper limit and small values of the argument $\alpha$ exist:
\begin{align}	
X\left(k,\alpha\right)&\simeq -\sqrt{\frac{i}{2\uppi}}\frac{e^{\left[ik\left(1-\cos\alpha\right)\right]}}{2\sqrt{k}\sin\frac{\alpha}{2}}~,~~~~~\sqrt{k}|\sin\frac{\alpha}{2}|\gg 1 \label{eq:26}\\
X\left(k,\alpha\right)&\simeq \frac{1}{2}-\sqrt{\frac{k}{2\uppi i}}\alpha~,~~~~~~~~~~~~~~~\alpha \rightarrow 0 \label{eq:27}
\end{align}

At this point, it is helpful to introduce variable: $\upzeta = \upxi - \uptheta_{2}$ (see Fig.~\ref{fig:3}). Substituting to (\ref{eq:21}) and after some lengthy but otherwise basic algebraic manipulations, the integral representation for the scatered electric field takes the form:

\begin{equation}\label{eq:28}
\underline{E}^{R}=\dfrac{pk_{01}^{3}}{2\upvarepsilon_{0}\upvarepsilon_{1}}\sqrt{\dfrac{-2i}{\uppi k_{01}\uprho}}\cdot e^{ik_{01}r_{2}\cos\upzeta_{p}}\cdot\frac{1}{4\uppi i}\int_{S_{z}}Q\left(\upzeta+\uptheta_{2}\right)\cdot\frac{e^{ik_{01}r_{2}\left(\cos\upzeta-\cos\upzeta_{p}\right)}}{\sin\frac{\upzeta-\upzeta_{p}}{2}}~d\upzeta
\end{equation}
with $S_{z}$ now being the contour of integration in the $\upzeta$-plane (right plot of Fig.~\ref{fig:3}),  resulting from the above variable change (i.e. left shifted by $\uptheta_{2}$) and:
\begin{equation}\label{eq:29}
Q\left(\upzeta\right)=\hat{e}_{\uptheta}\left(\upzeta\right)\sin^{^{\frac{3}{2}}}\hspace{-1mm}\upzeta\cdot R_{\parallel}\left(\upzeta\right)\sin\frac{\upzeta-\upxi_{p}}{2}
\end{equation}

Then, the \emph{Residue Theory} \cite{20}, together with the \emph{Saddle Point Method} \cite{17,18} for (\ref{eq:28}), are utilized. As shown in Appendix \ref{App:D}, this approach leads to an expression for $\underline{E}^{R}$, in which special function $X\left(k,a\right)$ plays a crucial role. The result, in which $\upzeta_{p}=\upxi_{p}-\uptheta_{2}$ is the equivalent pole to the $\upzeta$-plane, is given by:

\begin{equation}\label{eq:30}
\underline{E}^{R}=-\hat{e}_{\uptheta_{2}}\dfrac{pk_{01}^{3}}{2\upvarepsilon_{0}\upvarepsilon_{1}}\sqrt{\dfrac{-2i}{\uppi k_{01}\uprho}}\cdot e^{ik_{01}r_{2}\cos\upzeta_{p}}\cdot\sin^{^{\frac{3}{2}}}\hspace{-1mm}\uptheta_{2}\sin\frac{\upzeta_{p}}{2}R_{\parallel}\left(\uptheta_{2}\right)X\left(k_{01}r_{2}, -\upzeta_{p}\right)
\end{equation}

The analysis for the magnetic field is totally analogous and yields the well-known relation between the $\underline{E}$--$\underline{H} $ constituents of the scattered EM field:

\begin{equation}\label{eq:31}
\underline{H}^{R} = -\hat{e}_{a}\frac{1}{Z_{1}}\vert\underline{E}^{R}\vert
\end{equation}

\noindent Expression (\ref{eq:31}) is an essential validation for the mathematical spectral domain approach followed throughout this paper.

\subsubsection{Non-near ground level asymptotic expression}
As discussed in Appendix \ref{App:D}, (\ref{eq:30}) is valid only when $k_{01}r_{2}$ is large, i.e. it describes the far field behavior (which is what matters for most telecommunications applications). As a result, if it is also true that $|\sin\frac{\upzeta_{p}}{2}|$ is not very small, then the asymptotic conditions of (\ref{eq:26}) are met and $X\left(k_{01} r_{2}, -\upzeta_{p}\right)$ may be approximated by $\displaystyle X\left(k_{01} r_{2}, -\upzeta_{p}\right)\simeq\sqrt{\frac{i}{2\uppi}}\frac{e^{\left[ik_{01}r_{2}\left(1-\cos\upzeta_{p}\right)\right]}}{2\sqrt{k_{01}r_{2}}\sin\frac{\upzeta_{p}}{2}}$. Since, $\upxi_{p}\simeq\uppi/2$, the condition that satisfies the requirement that $\sin\frac{\upzeta_{p}}{2}$ does not approach zero, is equivalent to $\uppi/2 - \uptheta_{2} > \upvarepsilon$, with $\upvarepsilon$ sufficiently large. For literally large values of $k_{01}r_{2}$, it is just enough to require that $\uptheta_{2}\ne\pi/2$, such that the condition of (\ref{eq:26}), namely $\sqrt{k_{01}r_{2}}\hspace{-1mm}\cdot \left.\hspace{-1mm}\middle|\sin\left(\frac{\uptheta_{2}}{2}-\frac{\uppi}{4}\right)\hspace{-1mm}\middle|\right. \hspace{-2mm}\gg\hspace{-1mm} 1$, is met. The reflected field will then be given by\footnotemark : 

\footnotetext{Based on the geometry of Fig.~\ref{fig:1}, $\uptheta_{2}\in\left[0, \uppi/2\right]$}

\begin{equation}\label{eq:32}
\underline{E}^{R}\simeq-\hat{e}_{\uptheta_{2}}\cdot R_{\parallel}\left(\uptheta_{2}\right)\frac{pk_{01}^{2}}{4\uppi\upvarepsilon_{0}\upvarepsilon_{1}r_{2}}\cdot\sin\uptheta_{2}\cdot e^{ik_{01}r_{2}},~~~~~\uptheta_{2}<\frac{\uppi}{2}
\end{equation}

\noindent which is essentially the far field formula of a dipole source, located at the specular point A\'{} of Fig.~\ref{fig:1}, multiplied by the Fresnel reflection coefficient, $R_\parallel\left(\uptheta_{2}\right)$.

\subsubsection{Near ground level asymptotic expression}
Asymptotic expressions (\ref{eq:17}) and (\ref{eq:31}) for the direct and scattered field respectively, represent outgoing spherical waves. Now the conditions for the generation of the, well-known in the literature, \emph{surface waves} are studied \cite{1,2,3,4,5,6,19}. As their name imply, these waves are constrained close to the the planar interface, so a reasonable choice is to examine the field's behavior at sliding observation angles, ie at $\uptheta_{2}\rightarrow\uppi/2$.

In such case, it can be argued that $\hat{e}_{\uptheta_{2}} \simeq -\hat{e}_{x}$. Moreover, from (\ref{eq:20}), it is also valid to state that $R_{\parallel}\left(\uptheta_{2}\right)\simeq -1 $ and finally, since $\upzeta_{p}\rightarrow 0$, special function $X$ may be approximated by (\ref{eq:27}). Detailed analysis for reaching an asymptotic expression for $\underline{E}^{R}$ is given in Appendix E, with the final result being:

\begin{equation}\label{eq:33}
\underline{E}^{R}_{s}\simeq\hat{e}_{x}\updelta\frac{pk_{01}^{3}}{4\upvarepsilon_{0}\upvarepsilon_{1}}\cdot\frac{1}{\sqrt{\uppi k_{01}\uprho}}\cdot e^{-\updelta k_{01}\left(x+x_{0}\right)}\cdot e^{i\left(k_{01}\uprho+\uppi/2\right)},~~~~~ \updelta = \sqrt{\frac{\upomega\upvarepsilon_{0}\upvarepsilon_{1}}{2\upsigma}} 
\end{equation}

\noindent with the index ``s'' in $\underline{E}^{R}_{s}$ denoting the surface wave characteristics of the resulting expression, (\ref{eq:33}).

\section{Conclusions and Future Research}\label{sec:concl}
Under assumption $\upsigma\gg\upomega\upvarepsilon_{0}$, described in Section \ref{sec:asym_expr}, (\ref{eq:30}) represents the expression for the scattered electric field and includes special function $X\left(k_{01}r_{2}, -\upzeta_{p}\right)$, a function which, as mentioned above, is related to well known Fresnel integrals and possesses useful properties. Therefore, together with (\ref{eq:17}), for the LOS field, they may be considered as the complete analytical solution for the Sommerfeld Radiation problem, for this particular case.

Further analysis of  (\ref{eq:30}), revealed useful insights as for the nature of the EM field. Far away from the ground - air interface, (\ref{eq:32}) holds and the scattered field takes the form of an outgoing spherical wave (which of course resembles a plane wave at very large distances from the source) and together with the LOS field of (\ref{eq:17}), form what is frequently called in the literature as ``space wave'' \cite{5,19}. On the contrary, at sliding observation angles, that is at observation points that are far away from the source and simultaneously very close to the ground (so as $\uptheta_{2}\simeq \uppi/2$), the space wave diminishes. This would also be a direct outcome of (\ref{eq:17}) and (\ref{eq:32}), since in this case $R_{\parallel}\left(\uptheta\right)\simeq -1$, $r_{1}\simeq r_{2} \text{ and } \uptheta_{1}\simeq\uptheta_{2}$. Actually, at sliding observation angles, the scattered field of (\ref{eq:30}) takes the asymptotic form of (\ref{eq:33}), which is essentially a cylindrical wave and has clear surface wave characteristics (i.e. constrained near the ground), described by the exponentially decaying factor, $e^{-\updelta k_{01}\left(x+x_{0}\right)}$. It decays slower that the direct field ($\sim r^{-1/2} ~\text{vs}~\sim r^{-1}$ respectively) and hence it is the prevailing field far away from the source (i.e. in the far field). Moreover, the location of the transmitting source also plays a major role to the determination of the surface wave. Due to the presence of $x_{0}$ in the exponentially decaying factor of (\ref{eq:33}), it is expected to diminish (always in an exponential manner) for high-altitude antenna sources. Finally, (\ref{eq:33}) succeeds to describe the field behavior for the extreme case where $\upsigma\rightarrow\infty$. In accordance to the literature and theory, it vanishes due to its dependence on factor $\updelta=\sqrt{\upomega\upvarepsilon_{0}\upvarepsilon_{1}/2\upsigma}$.

Future research will focus on further investigation of special function $X\left(k,a\right)$ and its properties, as well as other special functions that could be utilized in the evaluation of the EM field integral expression, namely (\ref{eq:19}). The goal is to provide useful asymptotics for every possible case, instead of only for the $\upsigma\gg\upomega\upvarepsilon_{0}$ one, considered here and also for the transmitted field, propagating below the ground level, given by (\ref{eq:11})--(\ref{eq:12}). Furthermore, comparisons between found asymptotic formulas against the numerical calculation of the respective integral formulas, (\ref{eq:9})--(\ref{eq:12}), will be examined. These comparisons will eventually orientate the limits of applicability of the proposed approximations. For this purpose, comparisons with the results of other research groups (including classical approximations, \cite{3,4,7,8,9,10,11,13}), as well as with experimental data, are also in the plans.

\

\appendixx{On the transformation of EM field impoproper integrals to contour integrals in the complex plane}\label{App:A}

\normalsize
Each of the expressions (\ref{eq:7})--(\ref{eq:8}) and (\ref{eq:9})--(\ref{eq:10}), for the direct and scattered EM field respectively, are improper integrals along the real infinite integration  axis, $\left(-\infty, +\infty\right)$. Since $k_{\uprho}$ is the integration variable, they take the following general form:

\begin{equation}\label{eq:A1}
I = \int_{-\infty}^{+\infty}f\left(k_{\uprho}\right)~dk_{\uprho}= \underbrace{\int_{-\infty}^{-k_{01}}f\left(k_{\uprho}\right)~dk_{\uprho}} +\underbrace{\int_{-k_{01}}^{+k_{01}}f\left(k_{\uprho}\right)~dk_{\uprho}} +\underbrace{\int_{+k_{01}}^{+\infty}f\left(k_{\uprho}\right)~dk_{\uprho}}
\end{equation}

\hspace{64mm}$I_{1}$\hspace{30mm}$I_{2}$\hspace{29mm}$I_{3}$

%\vspace{2mm}
For \large{\textit{I}\textsubscript{2}}, it is easy to set $k_{\uprho}=k_{01}\sin t \Rightarrow dk_{\uprho} = k_{01} \cos t ~dt$, with \textit{t} being a real dummy variable in the $\left[-\uppi/2, \uppi/2\right]$ interval. This transforms the integral to:

\begin{equation}\label{eq:A2}
I_{2}=\int_{-\frac{\uppi}{2}}^{+\frac{\uppi}{2}}f\left(k_{01}\sin t\right)k_{01}\cos t~dt
\end{equation}

In the complex $\upxi$-plane of Fig.~\ref{fig:3}, (\ref{eq:A2}) simply represents, the contour integral of complex function $F\left(\upxi\right)=f\left(k_{01}\sin\upxi\right)k_{01}\cos\upxi$, over the $\gamma_{2}$ segment of $S_{z}$. Indeed, the parametric expression of segment $\gamma_{2}$ of Fig.~\ref{fig:2} is simply $\upxi\left(t\right) = t$, with $t\in\left[-\uppi/2, \uppi/2\right]$. Hence by definition:

\begin{equation}\label{eq:A3}
\int_{\gamma_{2}}F\left(\upxi\right)~d\upxi = \int_{-\frac{\uppi}{2}}^{+\frac{\uppi}{2}}F\left(\upxi\left(t\right)\right)\upxi'\left(t\right)~dt=\int_{-\frac{\uppi}{2}}^{+\frac{\uppi}{2}}f\left(k_{01}\sin t\right)k_{01}\cos t~dt=I_{2} 
\end{equation}

Now consider integral \large{\textit{I}\textsubscript{3}}. A sufficient variable transformation that will effectively map to $\left[k_{01}, +\infty\right]$ is: $k_{\uprho}=k_{01}\cosh t \Rightarrow dk_{\uprho} = k_{01} \sinh t ~dt$. With this variable change, interval $\left[k_{01}, +\infty\right]$ is mapped to $\left[0, +\infty\right]$ and \large{\textit{I}\textsubscript{3}} becomes:

\begin{equation}\label{eq:A4}
I_{3}=\int_{0}^{\infty}f\left(k_{01}\cosh t\right)k_{01}\sinh t~dt=\int_{0}^{\infty}f\left[k_{01}\sin\left(\frac{\uppi}{2}-it\right)\right]k_{01}\cos\left(\frac{\uppi}{2}-it\right)\left(-i\right)~dt
\end{equation}

\noindent for which well known relations of hyperbolic and trigonometric functions have been utilized, namely: $\cosh t=\cos it=sin\left(\uppi/2 - it\right)$, $\sinh t=-i\sin it=-i\cos\left(\uppi/2 - it\right)$ As was the case for \large{\textit{I}\textsubscript{2}}, (\ref{eq:A4}) expresses the contour integral of $F\left(\upxi\right)$, this time, over segment $\gamma_{3}$ of $S_{z}$. To show this, consider the parametric expression of $\gamma_{3}$, which is $\upxi\left(t\right) = (\uppi/2)-it$, $t\in\left[0, \infty\right]$. Hence, the contour integral of $F\left(\upxi\right)$ over $\gamma_{3}$ is:
\begin{equation}\label{eq:A5}
\int_{\gamma_{3}}F\left(\upxi\right)d\upxi = \int_{0}^{\infty}\hspace{-2mm}F\left(\upxi\left(t\right)\right)\upxi'\left(t\right)dt=\int_{0}^{\infty}\hspace{-2mm}f\left[k_{01}\sin\left(\frac{\uppi}{2}-it\right)\right]k_{01}\cos\left(\frac{\uppi}{2}-it\right)\left(-i\right)dt=I_{3}
\end{equation}

Similarly, it is easily derived that:

\begin{equation}\label{eq:A6}
I_{1}=\int_{\gamma_{1}}F\left(\upxi\right)~d\upxi 
\end{equation}

Overall, from (\ref{eq:A3}), (\ref{eq:A5}), (\ref{eq:A6}) and Fig.~\ref{fig:2}  ($S_{z}\equiv\gamma_{1}+\gamma_{2}+\gamma_{3}$), expression (\ref{eq:A1}) becomes:

\begin{equation}\label{eq:A7}
I = \int_{-\infty}^{+\infty}f\left(k_{\uprho}\right)~dk_{\uprho}= \int_{\gamma_{1}}F\left(\upxi\right)~d\upxi+\int_{\gamma_{2}}F\left(\upxi\right)~d\upxi+\int_{\gamma_{3}}F\left(\upxi\right)~d\upxi=\int_{S_{z}}F\left(\upxi\right)~d\upxi
\end{equation}

%\subsection{Sub-Appendix}

%\begin{equation}
%E=mc^2
%\end{equation}

\appendixx{Evaluating the position of the pole}\label{App:B}
\normalsize
For $\upsigma\gg\upomega\upvarepsilon_{0}$, (\ref{eq:22}) can be written as:

\begin{equation}\label{eq:B1}
\cos\upxi_{\text{p}}=-\sqrt{\frac{\upvarepsilon_{1}}{\upvarepsilon_{1}+\upvarepsilon_{2}}}=\sin\left(\frac{\pi}{2}-\upxi_{\text{p}}\right)\simeq\frac{\pi}{2}-\upxi_{\text{p}}
\end{equation}

\noindent where the approximation for the sin() is justified by the small magnitude of complex number $\sqrt{\frac{\upvarepsilon_{1}}{\upvarepsilon_{1}+\upvarepsilon_{2}}}$ (again it is assumed here that $\upvarepsilon_{2}\equiv\dot{\upvarepsilon}_{2} = \upvarepsilon_{2} + i\frac{\upsigma}{\upomega\upvarepsilon_{0}}$ with the latter $\upvarepsilon_{2}$ being real):
 
\begin{equation}\label{eq:B2}
\sqrt{\frac{\upvarepsilon_{1}}{\upvarepsilon_{1}+\upvarepsilon_{2}}}\equiv\sqrt{\frac{\upvarepsilon_{1}}{\upvarepsilon_{1}+\upvarepsilon_{2} + i\frac{\upsigma}{\upomega\upvarepsilon_{0}}}}=\sqrt{\frac{\upomega\upvarepsilon_{0}\upvarepsilon_{1}}{\upsigma}}\sqrt{\frac{1}{i+\frac{\upomega\upvarepsilon_{0}\left(\upvarepsilon_{1}+\upvarepsilon_{2}\right)}{\upsigma}}}=\sqrt{\frac{\upomega\upvarepsilon_{0}\upvarepsilon_{1}}{\upsigma}}\sqrt{\frac{1}{i+x}}
\end{equation} 

\noindent with $\displaystyle x=\frac{\upomega\upvarepsilon_{0}\left(\upvarepsilon_{1}+\upvarepsilon_{2}\right)}{\upsigma}\ll1$, for $\upsigma\gg\upomega\upvarepsilon_{0}$ Then, taking a MacLaurin series expansion for $f(x)=\sqrt{i+x}$ and keeping upto first order terms, it holds true that: 

\begin{equation}\label{eg:B3}
f\left(x\right)=\sqrt{i+x}=f\left(0\right)+f'\left(0\right)x+o\left(x^{2}\right)\simeq\sqrt{i}+\frac{1}{2\sqrt{i}}\cdot x=\frac{i+y}{\sqrt{i}}
\end{equation}

\noindent with $\displaystyle y=\frac{x}{2}=\frac{\upomega\upvarepsilon_{0}\left(\upvarepsilon_{1}+\upvarepsilon_{1}\right)}{2\upsigma}$. Substituting to (\ref{eq:B2}), the following is obtained:

\begin{equation}\label{eq:B4}
\sqrt{\frac{\upvarepsilon_{1}}{\upvarepsilon_{1}+\upvarepsilon_{2}}}=\sqrt{\frac{\upomega\upvarepsilon_{0}\upvarepsilon_{1}}{\upsigma}}\frac{\sqrt{i}}{y+i}=\sqrt{\frac{\upomega\upvarepsilon_{0}\upvarepsilon_{1}}{2\upsigma}}\cdot\frac{1+i}{y+i} 
\end{equation}

Finally, expressing $\displaystyle \frac{1+i}{y+i}$ to the ordinary complex form of $a+ib$, it is easy to get that:

\begin{equation}\label{eq:B5}
\frac{1+i}{y+i} = \frac{1+y}{1+y^{2}}+i\left(\frac{y^{2}+y}{1+y^{2}}-1\right)\stackrel{y^{2}\rightarrow 0}{\simeq} 1+y+i\left(y-1\right)\hspace{-1mm}=\hspace{-1mm} 1+\frac{\upomega\upvarepsilon_{0}\left(\upvarepsilon_{1}+\upvarepsilon_{1}\right)}{2\upsigma}+ i\left[\frac{\upomega\upvarepsilon_{0}\left(\upvarepsilon_{1}+\upvarepsilon_{1}\right)}{2\upsigma}-1\right]
\end{equation}

Substituting (\ref{eq:B5}) to (\ref{eq:B4}) and using (\ref{eq:B1}), we reach (\ref{eq:23}).

\appendixx{Associating the Etalon Integral to Probability integrals}\label{App:C}
\normalsize
Special function $X\left(k,\alpha\right)$ is given by \cite{24}:

\begin{equation}\label{eq:C1}
X\left(k,\alpha\right)=\frac{e^{-i\frac{\uppi}{4}}}{\sqrt{2\uppi}}\int_{\infty\sin\frac{\alpha}{2}}^{2\sqrt{k}\sin\frac{\alpha}{2}}\large{e^{\frac{it^{2}}{2}}~dt}
\end{equation}

For $\alpha\in[-\uppi, +\uppi]$, it is true that sgn($\alpha$)=sgn($\sin\frac{\alpha}{2}$). For the case where $\alpha>0$, the down limit of $X\left(k,\alpha\right)$ is $\displaystyle\infty\cdot\sin\frac{\alpha}{2}\rightarrow\infty$. Then by setting $t=\sqrt{2i}y$ and using the definition for the complementary error function\footnotemark, (\ref{eq:C1}), becomes:

\begin{equation}\label{eq:C2}
X\left(k,\alpha\right)=\frac{e^{-i\frac{\uppi}{4}}}{\sqrt{2\uppi}}\int_{\infty}^{\sqrt{-2ik}\sin\frac{\alpha}{2}}e^{-y^{2}}\cdot\sqrt{2i}~dy=\frac{1}{2}\text{erfc}\left[\sqrt{-2ik}\sin\frac{\alpha}{2}\right]
\end{equation} 

\footnotetext{$\text{erfc}\left(x\right)=\frac{2}{\sqrt{\uppi}}\int_{x}^{\infty}e^{-y^{2}}~dy$}

Similarly, for $\alpha<0$, $\displaystyle\infty\cdot\sin\frac{\alpha}{2}$ maps to $-\infty$. Setting again $t=\sqrt{2i}y$ to (\ref{eq:C1}), gives:

\begin{equation}\label{eq:C3}
X\left(k,\alpha\right)=\frac{e^{-i\frac{\uppi}{4}}}{\sqrt{2\uppi}}\int^{\infty}_{-\sqrt{-2ik}\sin\frac{\alpha}{2}}e^{-y^{2}}\cdot\sqrt{2i}~dy=-\frac{1}{2}\text{erfc}\left[-\sqrt{-2ik}\sin\frac{\alpha}{2}\right]
\end{equation} 

The combination of (\ref{eq:C1}) and (\ref{eq:C3}) for both cases $\alpha>0$ and $\alpha<0$, yields (\ref{eq:25}) 

\appendixx{On the contour integration over $S_{z}$}\label{App:D}
\normalsize
First, note that the integrand in (\ref{eq:28}) includes a phase factor, $e^{ik_{01}r_{2}\left(\cos\upzeta-\cos\upzeta_{p}\right)}$. Then on the assumption that $k_{01}r_{2}$ is a large parameter, several asymptotic methods for evaluating the integral exist, which all make the meaningful assumption that the main contribution to the integral's value comes from a small area, in the vicinity of a ``stationary point'' \cite{17,18}. The saddle point method to be used here is such a method. The conclusion is that for $k_{01}r_{2}\gg 1$, curves $S_{z}$ and $S$ may well be considered as part of the closed curve, shown in Fig.~\ref{fig:3}, in which the contribution of the dashed-line segments is neglectible, for being away from this stationary, or ``saddle point'', to be calculated below. Then by means of the Residue Theory and since no pole exists inside the aforementioned closed contour, it holds true:

\begin{equation}\label{eq:D1}
\int_{S_{z}}\hspace{-1mm} Q\left(\upzeta+\uptheta_{2}\right)\cdot\frac{e^{ik_{01}r_{2}\left(\cos\upzeta-\cos\upzeta_{p}\right)}}{\sin\frac{\upzeta-\upzeta_{p}}{2}}~d\upzeta = \int_{S}\hspace{-1mm} Q\left(\upzeta+\uptheta_{2}\right)\cdot\frac{e^{ik_{01}r_{2}\left(\cos\upzeta-\cos\upzeta_{p}\right)}}{\sin\frac{\upzeta-\upzeta_{p}}{2}}~d\upzeta
\end{equation}

As mentioned above, in order to evaluate $\displaystyle \int_{S} Q\left(\upzeta+\uptheta_{2}\right)\cdot\frac{e^{ik_{01}r_{2}\left(\cos\upzeta-\cos\upzeta_{p}\right)}}{\sin\frac{\upzeta-\upzeta_{p}}{2}}~d\upzeta$, the saddle point method is applied, under the precondition that $k_{01}r_{2}$ is a large parameter. It is easy to find that $\upzeta=0$ is the saddle point (found by solving $\dfrac{\partial}{\partial\upzeta}\left(\cos\upzeta - \cos\upzeta_{p}\right)=0)$. Hence, the integral may be evaluated as:

\begin{align}\label{eq:D2}
\hspace{-4mm}\int_{S}\hspace{-1mm} Q\left(\upzeta+\uptheta_{2}\right)\frac{e^{ik_{01}r_{2}\left(\cos\upzeta-\cos\upzeta_{p}\right)}}{\sin\frac{\upzeta-\upzeta_{p}}{2}}~d\upzeta = Q\left(\uptheta_{2}\right)\cdot\int_{S}\frac{e^{ik_{01}r_{2}\left(\cos\upzeta-\cos\upzeta_{p}\right)}}{\sin\frac{\upzeta-\upzeta_{p}}{2}}~d\upzeta=Q\left(\uptheta_{2}\right)\cdot X\left(k_{01}r_{2}, -\upzeta_{p}\right)
\end{align}

Substituting (\ref{eq:D2}) to (\ref{eq:28}) and also using (\ref{eq:29}) for $Q\left(\uptheta_{2}\right)$, expression (\ref{eq:30}) for $\underline{E}^{R}$ is obtained.

\appendixx{Asymptotics of the ``Surface Wave''}
Here, (\ref{eq:33}) is asymptotically derived from (\ref{eq:30}) in the limit $\uptheta_{2}\rightarrow\uppi/2$. The analysis is facilitated by defining auxiliary variable $\uptheta^{'}_{2}$, defined as:
\normalsize
\begin{equation}\label{eq:E1}
\uptheta^{'}_{2} = \uptheta_{2} - \updelta
\end{equation}
For $\updelta\rightarrow 0$, $\uptheta^{'}_{2}\rightarrow\uptheta_{2}$ and as a result from Fig.~\ref{fig:1} the following approximations may be made:

\begin{equation}\label{eq:E2}
\uprho\sin\uptheta^{'}_{2}\simeq\uprho,~~~~~\uprho\cos\uptheta^{'}_{2}\simeq\left(x+x_{0}\right) 
\end{equation}

It's useful to set small variable $\updelta$ equal to the infinitesimal quantity: $\sqrt{\frac{\upomega\upvarepsilon_{0}\upvarepsilon_{1}}{2\upsigma}}$ (valid since $\upsigma\gg\upomega\upvarepsilon_{0}$). Then (\ref{eq:23}) may be written as:

\begin{equation}\label{eq:E3}
\upxi_{\text{p}}\simeq \frac{\uppi}{2}+\updelta\left[1+\frac{\upvarepsilon_{1}+\upvarepsilon_{2}}{\upvarepsilon_{1}}\cdot\updelta^{2}-i\left(1-\frac{\upvarepsilon_{1}+\upvarepsilon_{2}}{\upvarepsilon{1}}\cdot\updelta^{2}\right)\right]\stackrel{\updelta^{2}\rightarrow 0}{\simeq}~~\frac{\uppi}{2}+\updelta-i\updelta
\end{equation}

\noindent As a result, the equivalent $\upzeta$-plane pole and its associated cosine are estimated as:

\begin{displaymath}
\upzeta_{p}=\upxi_{p}-\uptheta_{2} = \frac{\uppi}{2}-\uptheta^{'}_{2}-i\updelta
\end{displaymath}
\begin{equation}\label{eq:E4}
\cos\upzeta_{p} = \sin\left(\uptheta^{'}_{2}+i\updelta\right)=\sin\uptheta^{'}_{2}\cos i\updelta + \cos\uptheta^{'}_{2}\sin i\updelta\stackrel{\updelta\rightarrow 0}{\simeq}\sin\uptheta^{'}_{2}+i\updelta\cos\uptheta^{'}_{2}
\end{equation}

Now consider again (\ref{eq:30}):$~\underline{E}^{R}=-\hat{e}_{\uptheta_{2}}\dfrac{pk_{01}^{3}}{2\upvarepsilon_{0}\upvarepsilon_{1}}\sqrt{\dfrac{-2i}{\uppi k_{01}\uprho}}\cdot e^{ik_{01}r_{2}\cos\upzeta_{p}}\cdot\sin^{\frac{3}{2}}\uptheta_{2}\sin\frac{\upzeta_{p}}{2}R_{\parallel}\left(\uptheta_{2}\right)X\left(k_{01}r_{2}, -\upzeta_{p}\right)$

\noindent Since factors $\sin^{\frac{3}{2}}\uptheta_{2}$ and $\sin\frac{\upzeta_{p}}{2}$ appear only as amplitudes, they may be well estimated by setting $\uptheta_{2}\simeq \uppi/2$. Therefore, this time, adequate values for $\upzeta_{p}/2$ and its respective sine are:
 
\begin{displaymath}
\frac{\upzeta_{p}}{2}=\frac{\upxi_{p}-\uptheta_{2}}{2} \simeq \frac{\uppi}{4}+\updelta\frac{1-i}{2}-\frac{\uppi}{4}=\frac{\sqrt{2}}{2}e^{-i\uppi/4}\updelta
\end{displaymath}
\begin{equation}\label{eq:E5}
\sin\left(\frac{\upzeta{p}}{2}\right)\stackrel{\updelta\ll 1}{\simeq}\frac{\upzeta{p}}{2}=\frac{\updelta}{\sqrt{2}}\sqrt{-i}
\end{equation}

Finally, (\ref{eq:27}) is used for evaluating $X\left(k_{01}r_{2}, -\upzeta_{p}\right)$:

\begin{equation}\label{eq:E6}
X\left(k_{01}r_{2}, -\upzeta_{p}\right)\simeq\frac{1}{2}+\sqrt{\frac{k_{01}r_{2}}{2\uppi i}}\cdot\upzeta_{p}=\frac{1}{2}+\sqrt{\frac{k_{01}r_{2}}{2\uppi i}}\cdot\sqrt{-2i}\cdot\updelta
\end{equation}

Substituting, (\ref{eq:E2}), (\ref{eq:E4}), (\ref{eq:E5}) and (\ref{eq:E6}) to (\ref{eq:30}) and after neglecting 2nd order terms with respect to small parameter $\updelta$, we reach (\ref{eq:33}).

\end{document}